\documentstyle[12pt]{article}

\begin{document}

\setcounter{page}{1}

\centerline{\large {\bf A Model of Mesons based on $\chi$SB in the
Light-Front Frame}\footnote{Based on a lecture
given by L. Susskind at the workshop on `Theory of Hadrons and Light-Front QCD'
at Polona Zgorzelisko, Poland, August 15-25,1994.}}
\vskip.3in
\centerline{ L. Susskind}
\centerline{ Physics Department, Stanford University}
\centerline{ Stanford, CA 94309}
\vskip.2in
\centerline{M. Burkardt}
\centerline{Institute for Nuclear Theory, University of Washington}
\centerline{Seattle, WA 98195}
\vskip.6 in
\centerline{\bf Abstract}
\vskip.1in
Spontaneous breaking of chiral symmetry is discussed in the
light-cone framework. The essential ingredient is an infinite number
of constituents near zero light-cone momentum. These high (light-cone) energy
degrees of freedom freeze out and leave behind some explicit
symmetry breaking in the low (light-cone) energy effective Hamiltonian.
Connections with Regge theory and soft pion theorems are discussed.
Taking the order parameter to be the 4-th component of a chiral
4-vector, the effect of the spontaneous symmetry breaking
on meson masses and decay width is calculated and compared
with experimental data.
\vskip.1in

\vskip.3in
{\bf 1. Introduction}
\vskip.1in

Long before QCD, when I started thinking about field theory
in the light-cone (LC) frame [1], I was attracted by the fact that
in the LC frame the vacuum is just empty space and there is no
confusion between vacuum and particles. Since I was trying to
understand hadrons, it seemed unnecessary to spend time
understanding the vacuum --- and getting out ``absolutely nothing''.
Of course, I was wrong and the vacuum is very interesting, but
it still sounds like a nice idea to calculate hadrons in the
{\it  infinite momentum frame} [2], where everything moves so
fast relative to the vacuum that it would decouple from it in
some sense.

Around 1972, i.e. when QCD was considered as a candidate
for the strong interactions but before QCD became the
standard understanding of what
the strong interaction was, it became fairly clear that the basic
problems of strong interaction was twofold:
the confinement problem and the chiral symmetry breaking problem.
Since I was more interested in confinement, and since the
LC frame does not seem to be an especially helpful way to think
about confinement, I abandoned the LC and started thinking about
lattice gauge theory. But I also felt that chiral symmetry
breaking ($\chi$SB) is a problem that could very well be
understood in the LC frame. First it seems a bit of a puzzle how
a symmetry breaking can occur in the LC frame because, after all,
in the LC frame the vacuum is just the structureless
Fock space vacuum. In this lecture, I will provide a picture of
how $\chi$SB can occur in the LC framework and I will work out
some of the consequences.

The parton model is an interesting way to think about hadrons.
In the naive parton model, one simply pictures a fast moving
particle as being some collection of constituents with
relatively large momentum, such that when one boosts the
system, doubles its momentum, all these partons
double their momenta and so forth.
Upon boosting the system to infinite momentum
the partons would all become very far from $\eta=0$,
where $\eta$ is the fraction of the particle's longitudinal
momentum carried by the parton. Since all the vacuum activity
takes place at $\eta=0$, it seems very curious how these partons
(at finite $\eta$) could ``feel'' what is going on at $\eta=0$.

Chiral symmetry is the symmetry generated by $Q_5^\alpha$,
where $\alpha$ is some isospin index. In the LC frame,
this is just the sum over all partons of
the parton's helicity times its isospin times a plus or
a minus --- depending on whether it is a quark or an
antiquark
[5]. Therefore the conservation of chiral symmetry in the
LC frame is a very simple thing: it just says that some
kind of ``generalized helicity'' is conserved and it does
not appear to have to do with quark masses.
Since quark masses can be introduced into the LC frame
without causing problems about helicity conservation, one
encounters the next puzzle: how come that if one gives
a mass to a quark, chiral symmetry is broken in an
ordinary frame of reference but in the LC frame it is not?
Besides being confusing this is also a nuisance because
obviously one cannot break chiral symmetry by simply
adding an ``induced mass term'' to the LC Hamiltonian.

The right way to think about spontaneous breaking of
chiral symmetry on the LC is that it somehow manifests
itself through interactions between partons at finite
$\eta$ and $\eta=0$ (the vacuum).
The problem or puzzle with this is that matrixelements
connecting states which are separated by a large
distance in rapidity\footnote{
Rapidity is the logarithm of the LC-momentum $\eta$.}
are suppressed.
So ho how could the valence quarks possibly feel what
is going on at $\eta=0$?
After QCD was invented, there seemed to be a mechanism by
which things at finite $\eta$ can connect to $\eta=0$, namely
by emitting a soft gluon. Since the spin of gluons is larger
than that of quarks they can more easily connect regions
that are widely separated in rapidity.
However, the emission of a soft gluon does not flip the
helicity of the quark and therefore it does not seem that
gluons being emitted into the $\eta=0$ region has much to do
with the $\chi$SB.

Before we embark on constructing a model for spontaneous
symmetry breaking, I should emphasize some very fundamental
property of LC Hamiltonians: under a rescaling of the
LC momentum, $\eta \rightarrow \lambda \eta$,
the LC-Hamiltonian scales like $H \rightarrow H/\lambda$.
This looks like a dilatation symmetry if we think of the
$\eta$-axis as a spatial axis. Of course, it is not a spatial
axis, but if we assume that
things are short range coupled on the $\eta$-axis we can
consider it as if it were a spatial axis and formulate a
field theory on this axis.
The dilatation symmetry then reflects some underlying
scale invariance of this field theory and the right tool
for studying such a system is the renormalization group.

\vskip.3in
{\bf 2. The Long Arm of the Vacuum}\footnote{There is an expression in
English {\it the long arm of the law} which means you cannot
get away from the police, no matter how hard you try.}
\vskip.1in

A particle or hadron is not just a collection of partons
at finite $\eta$. In fact, partons fill the $\eta$-axis in a
way which gets denser and denser as one goes to smaller $\eta$:
according to Feynman and Bjorken, the number of constituents
per unit $\eta$ is given by $d\eta/\eta$.
The resulting accumulation of partons at low $\eta$ is usually
called the {\it wee partons}. The question here is: can one
imagine a process that would allow to transmit the information
about the $\chi$SB from $\eta=0$ through the chain of wee partons
to finite $\eta$; one at a time, so that at no place
large momentum is transferred.

At first this seems impossible: if one thinks of the
$\eta$-axis as being more or less short range coupled
(short ``range'' in $\eta$ space) then one can regard
the wee partons as some kind of $1+1$ dimensional system.
Since one cannot have spontaneous symmetry breaking of a
continuous symmetry in a $1+1$ dimensional system, this
seems forbidden right away. However, as the following model
demonstrates, there is in fact no problem about the
$1+1$ dimensionality. Let us consider a collection of
constituents which are distributed along the $\eta$-axis
according to $d\eta/\eta$. Each constituent is assumed to
have a LC momentum $\eta_i$ and some
internal degree of freedom. For simplicity,
we will consider a $U(1)$ symmetry here, which will not be
specified any further.
The $U(1)$ phase of each parton will
be denoted by $\phi_i$ with conjugate momentum $\Pi_{\phi_i}$.
The Hamiltonian is assumed to consist of a kinetic term,
$\Pi_{\phi_i}^2/\eta_i$, for each constituent plus
a nearest neighbor (in $\eta$-space) coupling of the
form $(\phi_i-\phi_{i-1})^2/\bar{\eta}$, where $\bar{\eta}$
can be chosen to be the average $\eta$ of the two constituents.
The $\eta$-dependence of these terms has been chosen such
that the LC-Hamiltonian scales like LC-momentum$^{-1}$.
One obtains
\begin{equation}
H= \sum_i \left(\frac{\Pi_{\phi_i}^2}{\eta_i}
+ g\frac{(\phi_i-\phi_{i-1})^2}{\eta_i+\eta_{i+1}}\right).
\end{equation}
The $d\eta/\eta$ distribution of partons can be built into
this model by furthermore assuming that $\Delta \eta$,
the distance between neighboring partons, goes like
$\varepsilon \eta$, where $\varepsilon$ is some small parameter,
characterizing the density of partons on the rapidity axis.
This Hamiltonian can be solved by a simple mathematical trick.
For small $\varepsilon$, one can make a continuum approximation,
and the Hamiltonian becomes
\begin{equation}
H = \int_0^1 d\eta \left( \Pi_{\phi_\eta}^2 + g\left( \nabla_\eta
\phi_\eta\right)^2 \right),
\label{eq:free}
\end{equation}
where we have arbitrarily picked $\eta=1$ as the endpoint of
the axis.
A nice order parameter for the $U(1)$ symmetry of this
system is $\cos \phi$.
Since Eq.(\ref{eq:free}) is just the Hamiltonian for a free
massive field, one finds that $\langle 0|\cos \phi |0\rangle$
is essentially given by
$\exp \left( -\langle 0|\phi^2 |0\rangle \right)$ (this result is
obtained by expanding $\cos \phi$ and calculating all the
contractions). Usually in quantum field theory, the
expectation value of the square of a field is infinite
(in this case it would be logarithmically infinite).
Therefore this kind of matrix element usually vanishes.
Of course this is an UV-divergence and the theory should be cut off.
A natural cutoff is provided by $\Delta \eta = \varepsilon \eta$,
the spacing between neighboring constituents before we made a
continuum approximation of the system and hence
$\varepsilon$ plays the role of some kind of UV-regulator here.
With such a cutoff in place one obtains
\begin{equation}
\langle 0|\phi_\eta^2 |0\rangle \sim \log \frac{\eta}{\Delta \eta}
=\log \frac{1}{\varepsilon},
\end{equation}
and as long as $\varepsilon$ is finite, i.e. as long as the density
of partons on the rapidity axis is not infinite, one gets a finite
nonvanishing matrix element for the order parameter
$\langle 0|\cos \phi |0\rangle$.
Why does this violate the usual rules about $1+1$ dimensions?
Usually, what happens is $1+1$ dimensional systems fluctuate
too strongly such that there is no order left.
Here the coupling, i.e. the spring constant between
neighboring partons, gets stronger and stronger as one approaches
$\eta=0$ so rapidly that the system is able to hold itself together
--- despite the fact that there is an infinite number of steps
between $\eta=0$ and finite $\eta$. It thus becomes possible
to have spontaneous symmetry breaking.

\vskip.3in
{\bf 3. Regge Theory}
\vskip.1in

Let us consider now the spatial distribution of the partons
in the transverse directions. In principle, there is some
wavefunction from which one can calculate this
distribution. However, many qualitative features can be
understood on the basis of very simple arguments.
First let us order the partons in sequence of rapidity.
We will again assume that things are short range coupled in
rapidity. Therefore, if we go down the chain in $\eta$,
the transverse position behaves on average like a random walk,
i.e. its square grows like the number of partons down the chain
\begin{equation}
r_\perp^2 \propto  -\log \eta.
\end{equation}
Assuming that the transverse spatial distribution at position $\eta$ is
Gaussian, $\rho({\vec r}^2_\perp,\eta) \sim
\exp(-{\vec r}^2_\perp/\log\eta)$, one can compute the
Fourier transform
\begin{equation}
F({\vec q}_\perp) \sim \exp(-{\vec q}^2_\perp \log \eta)
= \eta^{{-\vec q}^2_\perp}.
\label{eq:form}
\end{equation}
Of course, there are all kinds of dimensionful constants
which have been left out in these equations.

This formula (\ref{eq:form}) is a special case of a very
general feature, namely that many quantities have power law
dependence on $\eta$. The reason power laws keep
reoccuring is because of the abovementioned scale invariance.
The formal machinery which is usually being used to describe this
behavior is Regge pole theory.

As I have indicated already, other quantities are also likely
to be power law distributed. Let us consider
the average charge per parton. For small $\eta$ it has to
vanish --- otherwise the total charge carried by the partons
is infinite. Experimentally, one obtains for the average
charge per parton at position $\eta$ approximately
\begin{equation}
e(\eta) \approx \hat{e} \eta^{0.5}.
\end{equation}
$\hat{e}$ is called the {\it residue}. Now let us combine this
result with the transverse distribution of the partons to
calculate the electromagnetic form factor of the hadron.
One obtains
\begin{equation}
F^{EM}({\vec q}_\perp) = \int_0^1 \frac{d\eta}{\eta}
e(\eta) \eta^{-{\vec q}_\perp^2} =
\frac{\hat{e}}{{\vec q}_\perp^2+1/2},
\end{equation}
which represents a particle pole at $m^2=1/2$ and coupling
$\hat{e}$. Very generally, there is
a connection between these power laws and the
spectrum of particles coupling to the hadron.
So the two elements of the theory are
\begin{itemize}
\item that one can
think of the $\eta$-axis as an axis on which one can do field theory
and talk about distributions
\item that the
spatial distribution of partons satisfy certain assumptions, which allow one to
compute particle masses and couplings in terms of these
distributions.
\end{itemize}

It should be emphasized that the exponents usually do in
general not
depend on the hadron under consideration (the residues do very
well). Furthermore, the above discussion can be repeated for
off-diagonal matrix elements (i.e. transition amplitudes)
in which case the residues assume some kind of matrix structure,
i.e. they can be considered as operators acting between
hadronic states.

\vskip.3in
{\bf 4. $\chi$SB and the Pion}
\vskip.1in

As long as the rapidity axis is sufficiently short range
coupled, one can, in addition to densities on the
$\eta$-axis (above we have already discussed the charge density),
also introduce {\it currents} and, at least in the case
of conserved charges, there should also
be continuity equations, $\dot{\rho} + dj/d\eta=0$.
In the following we will apply these results to the
axial current, which can flow up and down the $\eta$-axis,
but which is more or less locally conserved. For this purpose,
let me introduce a chiral 4-vector
$\phi_\alpha$ on the $\eta$-axis. Note that $\alpha$ is NOT
a space time index but is related to $({\vec \pi},\sigma)$.
The 4-th component of $\phi_\alpha$, basically $\sigma$,
represents an order parameter for $\chi$SB.
These chiral 4-vectors tend to line up exactly as if they
were a ferromagnet. Therefore they will be called chiral
magnets in the following.
The point is that the couplings between all these
chiral magnets goes like $1/\eta$, i.e. the low $\eta$
degrees of freedom are extremely frozen. That is
characteristic of the LC-frame. Therefore, if we bring
in one ``external'' magnet, it will not be able to upset
this order. It will precess, like a spin in an external
magnetic field, and chiral charge will flow in and out
from the region near the external probe.

So let us assume now that there is a nonvanishing axial
charge density $j_5$ at small $\eta$ and demonstrate that
this implies a massless pion. Furthermore, we will see that
this massless pion couples to the axial current in a
particular way which allow one to make models in which one can
calculate properties of the pion and of the hadron spectrum.
First let us take the chirally symmetric Hamiltonian and add some small
term which explicitly breaks the chiral symmetry
\begin{equation}
H=H_{symm} + c \int_0^1 \frac{d\eta}{\eta} \phi_4(\eta),
\end{equation}
where $c$ is some small constant. A straightforward calculation
yields
\begin{equation}
-i\dot{Q}_{\alpha 5} =\left[H,Q_{\alpha 5}\right]=
c\int_0^1 \frac{d\eta}{\eta^2} \phi_\alpha(\eta),
\end{equation}
where $\alpha$ runs from 1 to 3.

In order to derive some observable consequences of the above
picture, let us assume that the matrix elements of
$\phi_\alpha$ are also power law behaved, i.e.
\begin{equation}
\phi_\alpha(\eta) \sim {\hat \phi}_\alpha \eta^{1+\mu},
\end{equation}
where $\mu$ is some number. This yields
\begin{equation}
-i\dot{Q}_{\alpha 5} = c\hat{\phi} \int_0^1 \frac{d\eta}{\eta^2}
\eta^{1+\mu} = \frac{c\hat{\phi}}{\mu}.
\end{equation}
The only way to keep $\dot{Q}_{\alpha 5} $ from vanishing when $c\rightarrow
0$,
is to let $\mu \rightarrow 0$ linearly at the same time.
In order to avoid having too many constants in this discussion,
let me just assume $\mu=c$. The next step is to use
the knowledge of how $\phi$ has to behave to compute
a formfactor for something which couples to $\phi_\alpha$.
The calculation more or less parallels the calculation
of the electromagnetic case above except that the
$1/2$ gets replaced by $c$ and one obtains a formfactor which
goes like $\hat{\phi}/({\vec q}_\perp^2 +c)$, indicating a particle pole
at $c$. Due to the (small) explicit symmetry breaking,
the pion has a mass $c$ and it becomes massless in the
``chiral'' limit ($c \rightarrow 0$). Furthermore we found that
the matrix element of $\dot{Q}_{\alpha 5}$ is, up to some
numerical constants, equal to the emission amplitude of a pion.

So far we have established that within a parton model with short
range couplings on the rapidity axis one can define, chiral
charges and order parameter. We have furthermore obtained a basic
equation for the on shell pion emission amplitude
\footnote{Here we put in some numbers which have been omitted above.}
\begin{equation}
T^\pi_{AB} =\frac{2}{if_\pi} \langle A| \dot{Q}_{\alpha 5} |B\rangle
= -\frac{2}{f_\pi} \left( M_A^2-M_B^2 \right) \langle A| Q_{\alpha 5} |B\rangle
,
\end{equation}
which is a generalization of the Goldberger-Treiman relation.

\vskip.3in
{\bf 5. Effective Hamiltonians}
\vskip.1in

On the $\eta$-axis we have degrees of freedom, which get more
and more strongly coupled as go down towards $\eta=0$.
Usually in physics when we encounter problems with higher
and higher energy scales, we cut off the theory and make an
effective theory. Instead of studying the whole chain,
we will try cut off the highest frequency parts, by introducing
a cutoff $\varepsilon$ on $\eta$. In the presence of the cutoff,
the scale invariance mentioned in section 2 is broken.
However, we still want the physics to be independent of the
artificial parameter $\varepsilon$. Therefore, when we construct
the effective Hamiltonian, we have to look for a consistency
of descriptions as we move the cutoff further and further away,
while keeping the physics fixed. In other words, what we have
to look for is an UV fixed point for the renormalization
group transformations of this system.

The Hamiltonian evidently consists of three pieces in general.
One of them, $H_{>\varepsilon}$, will have to do with the
degrees of freedom above the cutoff. The of course there will
be a part $H_{<\varepsilon}$. Furthermore, there will be a term
$H_\varepsilon$ which couples the degrees of freedom in these
two regions.
First of all, the degrees of freedom governed by
$H_{<\varepsilon}$ are high energy degrees of freedom
and it costs a large amount of energy to disrupt them.
\footnote{Note that they are UV only in the sense that in the
LC frame it costs a lot of energy to disrupt them.}
Therefore $H_{<\varepsilon}$ is effectively just a number.
Similarly, $H_{\varepsilon}$, to the extend that it depends on
degrees of freedom in the frozen region, one can also use vacuum
expectation values (VEVs). Of course for modes with $\eta > \varepsilon$
one must keep the full operator structure in $H_{\varepsilon}$.
Now since this system is frozen into a chirally asymmetric
configuration, it is very natural to assume that in replacing
the $\eta < \varepsilon$ modes by their VEV in $H_{\varepsilon}$,
one breaks the chiral symmetry of the rest of the chain.
For example, since the order parameter is a chiral 4-vector,
the entrance of the chiral 4-vector into the Hamiltonian should
be times another chiral 4-vector. Therefore we might expect
that $H_{\varepsilon}$ should be replaced by something which is just
the 4-th component of a chiral 4-vector associated with the degrees
of freedom at the new end of the chain, i.e. at $\eta=\varepsilon$.
Although the whole chain is chirally symmetric, in the process of
integrating out the degrees of freedom in the small  $\eta$ region,
an explicit chiral symmetry breaking for the rest of the chain has
been introduced. The situation is very similar to an atomic impurity
introduced into a ferromagnet. As a whole, the system is still
rotationally symmetric. However, for most practical purposes the
ferromagnet (without the impurity) can be regarded as frozen
and the effective Hamiltonian for the impurity is not
rotationally symmetric. The rotational multiplets are split and the
spin of the atom precesses in the ``external'' field provided
by the ferromagnet, thereby radiating off magnons. The time derivative of
the angular momentum of the atom is equal to the amplitude for emitting
a spin wave.

Now let us consider a specific scheme. What objects can one make out
of quarks which transform like a chiral 4-vector. In order to simplify
search for appropriate operators, it is useful to introduce the
$SU_2\times SU_2$ ``gamma'' matrices (which have nothing to do with the
true gamma matrices in the spin sense).
\begin{equation}
\gamma_\alpha=
\left(\begin{array}{c} 0 \quad \tau_\alpha \\ \tau_\alpha \quad 0 \end{array}
\right)
\quad \quad \quad \quad \quad\quad
\gamma_0=
\left(\begin{array}{c} 0 \quad i \\ -i \quad 0 \end{array}
\right)
\end{equation}
and $\gamma_5=\pm \sigma_z$, where the plus (minus) applies to
quarks (antiquarks). The spin-flavor spinor of a quark in this
basis has components $(u\uparrow,d\uparrow,u\downarrow,d\downarrow)$.
These gamma matrices
have the advantage that the indices here are chiral 4-vector indices.
The point is that possible objects for a one body operator,
which transform like a chiral 4-vector, such as
$\psi^\dagger \gamma_4 \gamma_\alpha \psi$ do not commute with
$\sigma_z$ and
of course the chiral order parameter should commute with angular
momentum. It is thus not sufficient to build the operator out
of spin and isospin operators only, but one has to introduce
some spin orbit coupling as well.\footnote{See also Ref.[4] for
a more detailed discussion of this point.}
One strange and peculiar fact when one goes to the LC frame is that
chiral 4-vectors cannot be made out of one body operators.
In the following, we will explicitly construct such an operator
for the case of a two body system --- a meson consisting
of a quark and an antiquark (with LC momenta $\eta_q$ and
$\eta_{\bar{q}}$ respectively).
The most simple chiral 4-vector one can write for this system
involves at least four angular momentum states:
$|+\rangle$, $|-\rangle$, $|S\rangle$ and $|A\rangle$ which have the
properties:
$L_z|+\rangle=+|+\rangle$,
$L_z|-\rangle=-|-\rangle$ and $L_z|S\rangle=L_z|A\rangle=0$.
$|S\rangle$ and $|A\rangle$ are symmetric and antisymmetric under
interchange of $\eta_q$ and $\eta_{\bar{q}}$.
Out of these operator one can now construct the 4-th component of
a chiral four vector
\begin{equation}
\phi_4 \equiv \left[\left( {\vec \sigma}_q -{\vec \sigma}_{\bar{q}} \right)
\times {\vec B} \right]_z,
\label{eq:explop}
\end{equation}
where
\begin{eqnarray}
B_- &=& |S\rangle \langle +|\quad +\quad |-\rangle \langle S|
\nonumber\\
B_+ &=& |S\rangle \langle -|\quad +\quad |+\rangle \langle S|
{}.
\end{eqnarray}
The point is, Eq.(\ref{eq:explop}) has spin orbit couplings between
the quarks. Even though one can write down other candidates for a
the 4-th component of a chiral 4-vector, all other candidates are more
complicated.

A number of physical consequences can be derived on the basis of
these results (see Ref.[3] for a detailed discussion).
For example, the (bare) pion should, together with its friend the
(bare) $\sigma$,
form a chiral 4-vector. Therefore they cannot be in an orbital
angular momentum zero state! They are in an orbital angular momentum one
state. This is quite surprising because one expects from the
nonrelativistic quark model that the pion is in an s-wave ---
but that is simply not true in a LC frame.
The (bare) $\rho$ and the (bare) $a_1$ form a chiral tensor. They can be in
an s-wave. If one now adds $\phi_4$ from above (\ref{eq:explop}),
one discovers that the $\pi$ and the $a_1$ mix, while not affecting
the $\rho$ and the $\sigma$. In a phenomenological model one can
then calculate the wavefunctions of these hadrons by adjusting
the symmetry breaking term to fit the hadron masses and one can then,
in terms of these wavefunctions, calculate transition amplitudes.
This has been done in Ref.[3] for the matrix elements of the chiral
charge, which can then be used to calculate decay amplitudes involving the
emission of pions.

\begin{table}
\caption{Decay width for various meson decays involving
$\pi$-emission.}
\begin{tabular}{lcl}
\hline
Decay&Computed width (MeV) [3]&Expt. width (MeV) [6]\\
\hline
$a_1(1260)\rightarrow \sigma \pi$&47& $<28$\\
$a_1(1260)\rightarrow \rho \pi$& 185 & $\stackrel{<}{\sim} 400$\\
$\rho \rightarrow \pi \pi$&130&125\\
$a_2(1320)\rightarrow \rho \pi$&24&16\\
$a_0(980)\rightarrow \eta \pi$&75&?\\
$f_1(1285)\rightarrow  a_0(980) \pi$&18& 11\,$\left[\mbox{ignoring
$a_0 \rightarrow K \bar{K}$}\right]$\\
$\sigma \rightarrow \pi \pi$&475& $\sim 500$\\
$f_2(1270) \rightarrow \pi \pi$&108&157\\
$b_1(1235) \rightarrow a_0(980)\pi$&23&?\\
\hline
\end{tabular}
\end{table}

The calculations were done in a very simple scheme in which one
takes just 2 partons into account. Everything else is frozen.
And the 2 partons simply interact with the rest of the
frozen system. As the results in Table 1 show, with a few exceptions,
one does pretty well --- even with a minimal structure for the operators.
In fact, considering that two quarks is a rather crude approximation,
it is rather surprising how well the results fit the data.
One can also use this scheme to understand some hadron masses.

The basic upshot one should get from these results is that one
should think of these systems in a renormalization group way,
where one first truncates the system to a small number of degrees
of freedom and introduces an explicit breaking. Then one calculates
physical observables and moves the cutoff back in a sequence
of approximations --- requiring that the physics remains invariant.
This can also be translated into the statement that one must be looking for
a fixed point of the renormalization group.

\vskip.3in
{\bf References}
\vskip.1in

1. \parbox[t]{14.5cm}{L. Susskind, Phys. Rev. {\bf 165}, 1535 (1968); see also
J. Kogut and L. Susskind, Phys. Rep. {\bf 8C}, 75 (1973),
and references therein.}

2. S. Weinberg, Phys. Rev. {\bf 150}, 1313 (1966).

3. A. Casher and L. Susskind, Phys. Lett. {\bf 44B}, 171 (1973).

4. A. Casher and L. Susskind, Phys. Rev. D{\bf 9}, 436 (1974).

5. K. Wilson et. al., Phys. Rev. D{\bf 49}, 6720 (1994).

6. Particle data group, Phys. Rev. D{\bf 50}, 1173 (1994).

\end{document}